\documentclass[aps,preprint]{revtex4}%
\usepackage{amsfonts}
\usepackage{amsmath}
\usepackage{amssymb}
\usepackage{graphicx}%
\begin{document}
\title[Superstatistics for transition intensities]{Superstatistical random-matrix-theory approach to transition intensities in
mixed systems}
\author{A. Y. Abul-Magd}
\affiliation{Department of Mathematics, Faculty of Science, Zagazig University, Zagazig, Egypt}
\keywords{Superstatistics, Random-matrix theory, Transition intensity}
\pacs{05.40.-a, 05.45.Mt, 03.65.-w, 02.30.Mv}

\begin{abstract}
We study the fluctuation properties of transition intensities applying a
recently proposed generalization of the random matrix theory, which is based
on Beck and Cohen's superstatistics. We obtain an analytic expression for the
distribution of the reduced transition probabilities that applies to systems
undergoing a transition out of chaos. The obtained distribution fits the
results of a previous nuclear shell model calculations for some
electromagnetic transitions that deviate from the Porter-Thomas distribution.
It agrees with the experimental reduced transition probabilities for the
$^{26}$Al nucleus better than the commonly used $\chi^{2}$ distribution.

\end{abstract}
\date[Date text]{\today}
\startpage{1}
\endpage{2}
\maketitle

\section{INTRODUCTION}

Random matrix theory (RMT) is believed to describe quantal systems whose
classical counterpart has a chaotic dynamics \cite{mehta,haake,guhr}. In RMT
the matrix elements of the Hamiltonian in some basis are replaced with random
numbers. The theory is based on two main assumptions: (i) the matrix elements
are independent identically-distributed random variables, and (ii) their
distribution is invariant under unitary transformations. These lead to a
Gaussian probability density distribution for the matrix elements, $P\left(
H\right)  \varpropto\exp\left[  -\eta\text{Tr}\left(  H^{\dagger}H\right)
\right]  $. With these assumptions, RMT presents a satisfactory description
for numerous chaotic systems. Agreement with RMT is now considered to be a
signature of chaos in the quantum system. For time-reversal-invariant systems,
the appropriate form of random matrix theory is the Gaussian orthogonal
ensemble (GOE) ; that is the form which will mainly be considered in this paper.

For most systems, however, the phase space is partitioned into regular and
chaotic domains. These systems are known as mixed systems. Attempts to
generalize RMT to describe such mixed systems are numerous; for a review
please see \cite{guhr}. Most of these attempts are based on constructing
ensembles of random matrices whose elements are independent but not
identically distributed. Thus, the resulting expressions are not invariant
under base transformation. The first work in this direction is due to
Rosenzweig and Porter \cite{rosen}. They model the Hamiltonian of the mixed
system by a superposition of a diagonal matrix of random elements having the
same variance and a matrix drawn from a GOE. Therefore, the variances of the
diagonal elements of the total Hamiltonian are different from those of the
off-diagonal ones, unlike the GOE Hamiltonian in which the variances of
diagonal elements are twice those of the off-diagonal ones. Hussein and Pato
\cite{hussein} used the maximum entropy principle to construct "deformed"
random-matrix ensembles by imposing different constraints for the diagonal and
off-diagonal elements. This approach has been successfully applied to the case
of metal-insulator transition \cite{hussein1}. A recent review of the deformed
ensemble is given in \cite{hussein2}. Ensembles of band random matrices, whose
entries are equal to zero outside a band of limited width along the principal
diagonal, have often been used to model mixed systems
\cite{casati,fyodorov,haake}.

The past decade has witnessed a considerable interest devoted to the possible
generalization of statistical mechanics. Much work in this direction followed
Tsallis seminal paper \cite{Ts1}. Tsallis introduced a non-extensive entropy,
which depends on a positive parameter$\ q$ known as the entropic index. The
standard Shannon entropy is recovered for $q$ = 1. Applications of the Tsallis
formalism covered a wide class of phenomena; for a review please see, e.g.
\cite{Ts2}. Recently, the formalism has been applied to include systems with
mixed regular-chaotic dynamics in the framework of RMT
\cite{evans,toscano,bertuola,nobre,abul1,abul2}. However, the constraints of
normalization and existence of an expectation value for Tr$\left(  H^{\dagger
}H\right)  $ set up an upper limit for the entropic index $q$ beyond which the
involved integrals diverge. This restricts the validity of the non-extensive
RMT to a limited range\ near the chaotic phase \cite{abul1,abul2}.

Another extension of statistical mechanics is provided by the formalism of
superstatistics (statistics of a statistics), recently proposed by Beck and
Cohen \cite{BC}. Superstatistics arises as weighted averages of ordinary
statistics (the Boltzmann factor) due to fluctuations of one or more intensive
parameter (e.g. the inverse temperature). It includes Tsallis' non-extensive
statistics, for $q\geq1$, as a special case in which the inverse temperature
has a $\chi^{2}$-distributions. This formalism has been applied to model a
mixed system within the framework of RMT in Ref. \cite{supst,sust}. The joint
matrix element distribution was represented as an average over $\exp\left[
-\eta\text{Tr}\left(  H^{\dagger}H\right)  \right]  $ with respect to the
parameter $\eta$. The different choices of parameter distribution, which had
been studied in Beck and Cohen's paper \cite{BC}, were considered in
\cite{supst}. The parameter distribution has also been estimated \cite{sust}
by applying the principle of maximum entropy, as done by Sattin \cite{sattin}.
Explicit analytical results were obtained for the level density and the
nearest neighbor-spacing distributions.

Matrix elements of \ transition operator probe the system's wave functions so
that their statistical fluctuations provide additional information. In chaotic
systems, the reduced transition probabilities follow the Porter-Thomas
distribution \cite{porter}. This is a $\chi^{2}$-distribution of one degree of
freedom. As the system becomes more regular, the transition probabilities
deviate from the Porte-Thomas distribution. To account for these deviations,
Alhassid and Novoselsky \cite{alhassid} suggested that the transition widths
in mixed system may be analyzed in terms of a $\chi^{2}$-distribution of a
lower degree of freedom. The latter distribution does not fit well the
empirical distributions but consists with the observed number of weak
transitions as compared with the Porter-Thomas distribution (see, e.g.
\cite{adams,barbosa,hamoudi}). The distributions of experimental reduced
transition probabilities $B$ in $^{26}$Al \cite{adamsx} and $^{30}$P
\cite{shriner} expressed as functions of the log$B$ have peaks at
log$B<0$\ while all the $\chi^{2}$\ distributions are peaked at log$B=0$. We
show in the present paper that the superstatistical RMT provides us with a
more suitable generalization of the Porter-Thomas distribution. In Section II
we briefly review the concept of superstatistics and the necessary
generalization required to express the characteristics of the spectrum of a
mixed system into an ensemble of chaotic spectra with different local mean
level density. The evolution of the reduced transition-intensity distribution
during the stochastic transition induced by increasing the local-density
fluctuations is considered in Section III. Section IV demonstrates the quality
of fit achieved by the obtained transition-intensity distribution by comparing
its prediction with the results of a shell-model calculation by Hamoudi et al.
\cite{hamoudi}. The conclusion of this work is formulated in Section 5.

\section{SUPERSTATISTICAL RMT}

To start with, we briefly review the superstatistics concept as introduced by
Beck and Cohen \cite{BC}. Consider a non-equilibrium system with
spatiotemporal fluctuations of the inverse temperature $\beta$. Locally, i.e.
in spatial regions (cells) where $\beta$ is approximately constant, the system
may be described by a canonical ensemble in which the distribution function is
given by the Boltzmann factor $e^{-\beta E}$, where $E$ is an effective energy
in each cell. In the long-term run, the system is described by an average over
the fluctuating $\beta$. The system is thus characterized by a convolution of
two statistics, and hence the name \textquotedblright
superstatistics\textquotedblright. One statistics is given by the Boltzmann
factor and the other one by the probability distribution $f(\beta)$ of $\beta$
in the various cells. One obtains Tsallis' statistics when $\beta$ has a
$\chi^{2}$ distribution, but this is not the only possible choice. Beck and
Cohen give several possible examples of functions which are possible
candidates for $f(\beta)$. Sattin \cite{sattin} suggested that, lacking any
further information, the most probable realization of $f(\beta)$ will be the
one that maximizes the Shannon entropy. Namely this version of superstatistics
formalism will now be applied to RMT.

\subsection{Joint distribution of the  matrix-elements}

Gaussian random-matrix ensembles have several common features with the
canonical ensembles. In RMT, the square of a matrix element plays the role of
energy of a molecule in a gas. When the matrix elements are statistically
identical, one expects them to become distributed as the Boltzmann's. One
obtains a Gaussian probability density distribution of the matrix elements
\begin{equation}
P_{G}\left(  H\right)  \varpropto\exp\left[  -\eta\text{Tr}\left(  H^{\dagger
}H\right)  \right]
\end{equation}
by extremizing the Shannon entropy \cite{mehta} subjected to the constraints
of normalization and existence of the expectation value of Tr$\left(
H^{\dagger}H\right)  $. Here Tr means trace and $H^{\dagger}$\ is the
Hermitian cojugate of $H$. The quantity\ Tr$\left(  H^{\dagger}H\right)
$\ plays the role of the effective energy of the system, while the role of the
inverse temperature $\beta$ is played by $\eta$, being twice the inverse of
the matrix-element variance.

Our main assumption is that Beck and Cohen's superstatistics provides a
suitable description for systems with mixed regular-chaotic dynamics. We
consider the spectrum of a mixed system as made up of many smaller cells that
are temporarily in a chaotic phase. Each cell is large enough to obey the
statistical requirements of RMT but has a different distribution parameter
$\eta$ associated with it, according to a probability density $\widetilde
{f}(\eta)$. Consequently, the superstatistical random-matrix ensemble that
describes the mixed system is a mixture of Gaussian ensembles. Its
matrix-element joint probability density distributions obtained by integrating
distributions of the form in Eq. (1) over all positive values of $\eta$\ with
a statistical weight $\widetilde{f}(\eta)$,
\begin{equation}
P(H)=\int_{0}^{\infty}\widetilde{f}(\eta)\frac{\exp\left[  -\eta
\text{Tr}\left(  H^{\dagger}H\right)  \right]  }{Z(\eta)}d\eta,
\end{equation}
where $Z(\eta)=\int\exp\left[  -\eta\text{Tr}\left(  H^{\dagger}H\right)
\right]  d\eta$. Here we use the \textquotedblright B type
superstatistics\textquotedblright\ \cite{BC}. The distribution in Eq. (2) is
isotropic in the matrix-element space. Relations analogous to Eq. (1) can also
be written for the joint distribution of eigenvalues as well as any other
statistic that is obtained from it by integration over some of the
eigenvalues, such as the nearest-neighbor-spacing distribution and the level
number variance. The distribution $\widetilde{f}(\eta)$ has to be
normalizable, to have at least a finite first moment and to be reduces a delta
function as the system becomes fully chaotic.

An analogous ensemble made of a superposition of random-matrix ensembles has
recently been considered by Muttalib and Klauber \cite{muttalib}. These
authors have been seeking for generalizations of Gaussian random-matrix
ensembles, with the probability distributions $P(H)$ that are functions of the
single variable Tr$\left(  H^{\dagger}H\right)  $ like the distribution (2)
that follows here from the concept of superstatistics. However, they work not
directly with the distributions $P(H)$ themselves, but with the associated
characteristic functions defined as the Fourier transforms
\begin{equation}
C(T)=\int\exp\left[  i\text{Tr}\left(  T^{\dagger}H\right)  \right]  P(H)dH.
\end{equation}
They prove, among other things, that if $C(T)$ is a function of Tr$\left(
T^{\dagger}T\right)  $ only, then the most general $C(T)$, valid for
random-matrix ensembles of arbitrarily large dimension can be represented as
\begin{equation}
C(T)=\int f(b)\exp\left[  -b\text{Tr}\left(  T^{\dagger}T\right)  \right]  db.
\end{equation}
The inverse Fourier transformation of $C(T)$\ then leads to an expression for
$P(H)$ similar to the one in Eq.\ (2). We consider their result as a
justification of using Eq. (2) for ensembles of matrices of dimensions
$N\rightarrow\infty$.

\subsection{Marginal distribution for a single matrix-element}

Unlike the Gaussian random-matrix ensembles, the superstatistical ensemble has
correlated matrix elements. This can clearly be seen by the fact that the
joint distribution function defined by Eq. (2) does not factorize into a
product of distributions of the individual matrix elements. However, it is not
difficult to obtain a marginal distribution for each of the individual matrix elements.

We shall confine our consideration for the GOE; the generalization to the
other symmetry universalities is straightforward. In this case,
\begin{equation}
Tr\left(  H^{\dagger}H\right)  =Tr(H^{2})=\sum_{k}H_{kk}^{2}+2\sum_{k<l}%
H_{kl}^{2}.
\end{equation}
Integrating the joint distribution $P(H)$ over all the matrix elements except
one, say $H_{if},$ we obtain%
\begin{equation}
p_{if}(H_{if})=\int_{0}^{\infty}\widetilde{f}(\eta)\sqrt{2\eta/\pi}\exp\left[
-2\eta H_{if}^{2}\right]  d\eta.
\end{equation}
The parameter $\eta$ essentially defines the energy scale of the individual
ensembles, whose superposition compose the superstatistical ensemble. We
therefore assume that the distribution (6) will hold for the superstatistical
distribution of any physical quantity having the dimension of energy, which is
represented as a Gaussian random variable in the case of GOE.

We note that the distribution function of superstatistical ensemble depends on
the matrix elements through Tr($H^{2})$ which is base invariant. In other
words, the fuction $P(H)$ is invariant under rotation in the spcae of matrix
elements. Therefore the distributions $p_{if}$ have the same form for all
off-diagonal matrix elements of $H$. 

\subsection{Parameter distribution}

Following Sattin \cite{sattin}, we use the principle of maximum entropy to
evaluate $\widetilde{f}(\eta)$. Lacking a detailed information about the
mechanism causing the deviation from the prediction of RMT, the most probable
realization of $\widetilde{f}(\eta)$ will be the one that extremizes the
Shannon entropy
\begin{equation}
S=-\int_{0}^{\infty}\widetilde{f}(\eta)\ln\widetilde{f}(\eta)d\eta
\end{equation}
with the following constraints:

\textbf{Constraint 1}. The major parameter of RMT is $\eta$ defined in Eq.
(1). Superstatistics was introduced in Eq. (2) by allowing $\eta$ to fluctuate
around a fixed mean value $\left\langle \eta\right\rangle $. This implies the
existence of this mean value%
\begin{equation}
\left\langle \eta\right\rangle =\int_{0}^{\infty}\widetilde{f}(\eta)\eta
d\eta.
\end{equation}

\textbf{Constraint 2}. The fluctuation properties are usually defined for
unfolded spectra, which have a unit mean level spacing. The mean level density
is proportional to the inverse square root of $\eta.$ We thus require the
existence of the integral
\begin{equation}
\int_{0}^{\infty}\widetilde{f}(\eta)\eta^{-1/2}d\eta=1.
\end{equation}

Therefore, the most probable $\widetilde{f}(\eta)$ extremizes the functional
\begin{equation}
F=-\int_{0}^{\infty}\widetilde{f}(\eta)\ln\widetilde{f}(\eta)d\eta-\lambda
_{1}\int_{0}^{\infty}\widetilde{f}(\eta)\eta d\eta-\lambda_{2}\int_{0}%
^{\infty}\widetilde{f}(\eta)\eta^{-1/2}d\eta,
\end{equation}
where $\lambda_{1}$ and $\lambda_{2}$ are Lagrange multipliers. As a result,
we obtain
\begin{equation}
\widetilde{f}(\eta)=C\exp\left[  -\alpha\left(  \frac{\eta}{\eta_{0}}+2\left(
\frac{\eta_{0}}{\eta}\right)  ^{1/2}\right)  \right]
\end{equation}
where $\alpha$ and $\eta_{0}$ are parameters, which can be expressed in terms
of the Lagrange multipliers $\lambda_{1}$ and $\lambda_{2}$, and $C$ is a
normalization constant. The latter is given by
\begin{equation}
C=\frac{\alpha\sqrt{\pi}}{\eta_{0}G_{03}^{30}\left(  \left.  \alpha
^{3}\right\vert 0,\frac{1}{2},1\right)  }.
\end{equation}
Here $G_{03}^{30}\left(  \left.  x\right\vert b_{1},b_{2},b_{2}\right)  $ is a
Meijer's G-function \cite{luke,wolfram}; see also the appendix of Ref.
\cite{sust}. The parameter distribution $\widetilde{f}(\eta)$ is peaked at
$\eta_{0}$ and tends to a delta function as $\alpha\rightarrow\infty.$ The
value of $\eta_{0}$ will be fixed in the next section while the parameter
$\alpha$ will be considered as the tuning parameter for the stochastic transition.

A parameter distribution analogous to $\widetilde{f}(\eta)$ is obtained in
\cite{sust}, where the variable $\eta$ is replaced by the local mean level
density. This distribution is used in \cite{sust} to derive expressions for
the level density distribution, the nearest-neighbor spacing distribution, and
the two-level correlation function for spectra of superstatistical ensembles.
The expression obtained for the level-density distribution has a finite value
at the center of the spectrum for arbitrarily large ensemble dimension $N$,
and thus satisfies a necessary condition on $\widetilde{f}(\eta),$ which is
required by Muttalib and Klauber \cite{muttalib}.

\section{Transition-intensity distribution}

The probability $B_{if}$ of a transition from the initial configuration
$\left\vert i\right\rangle $ to the final configuration $\left\vert
f\right\rangle $ is given by%
\begin{equation}
B_{if}=\left\vert W_{if}\right\vert ^{2},
\end{equation}
where%
\begin{equation}
W_{if}=\left\langle f\left\vert O\right\vert i\right\rangle
\end{equation}
is the square of the transition operator $O$ in a special basis. In a chaotic
system, the eigenstates $\left\vert i\right\rangle $ and $\left\vert
f\right\rangle $ are believed to be very complicated. If the operator $O$
conserves time reversibility, the matrix elements $W_{if}$ are real. For a
chaotic system, it is reasonable to assume that $W_{if}$\ are
identically-distributed Gaussian random variable. This entails that the
transition intensities can be represented by a random variable that takes the
values%
\begin{equation}
y_{if}=\frac{B_{if}}{\left\langle B_{if}\right\rangle }%
\end{equation}
where $\left\langle B_{ij}\right\rangle $ is a suitably defined local average
value \cite{adams}, and has a Porter-Thomas distribution%
\begin{equation}
P_{\text{PT}}(y)=\sqrt{\frac{\eta}{\pi y}}e^{-\eta y}.
\end{equation}
The parameter $\eta$ is defined by the requirement that $\left\langle
y\right\rangle =1,$ and is equal to 1/2. A more elaborate derivation of the
Porter-Thomas distribution is given by Barbosa et al. \cite{barbosa}.

We now derive the superstatistical generalization of the Porter-Thomas
distribution. For this purpose, we assume that the matrix elements $W_{if}%
$\ are distributed according to Eq. (6). The parameter $\eta$ in Eq. (16) is
no more considered as a constant but allowed to fluctuate according to the
distribution $\widetilde{f}(\eta)$. The superstatistical transition intensity
distribution is then given by%
\begin{equation}
P_{\text{Superstatistical}}(y)=\int_{0}^{\infty}\widetilde{f}(\eta)\sqrt
{\frac{\eta}{\pi y}}e^{-\eta y}d\eta
\end{equation}
Substituting Eq. (11) for $\widetilde{f}(\eta)$ and integrating over $\eta,$
we obtain
\begin{equation}
P_{\text{Superstatistical}}(y)=\frac{\alpha}{\sqrt{\pi y}}\frac{G_{03}%
^{30}\left(  \left.  \alpha^{2}(a+\eta_{0}y)\right\vert 0,\frac{1}{2},\frac
{3}{2}\right)  }{\eta_{0}(y+a/\eta_{0})^{3/2}G_{03}^{30}\left(  \left.
\alpha^{3}\right\vert 0,\frac{1}{2},1\right)  }%
\end{equation}
The parameter $\eta_{0}$ is determined from the requirement that $\left\langle
y\right\rangle =1$, which yields%
\begin{equation}
\eta_{0}=\frac{\alpha}{2}\frac{G_{03}^{30}\left(  \left.  \alpha
^{3}\right\vert 0,0,\frac{1}{2}\right)  }{G_{03}^{30}\left(  \left.
\alpha^{3}\right\vert 0,\frac{1}{2},1\right)  }.
\end{equation}
Replacing Meijer's G-function by its large-argument asymptotic expression%

\begin{equation}
G_{0,3}^{3,0}\left(  z\left\vert b_{1},b_{2},b_{3}\right.  \right)  \sim
\frac{2\pi}{\sqrt{3}}z^{(b_{1}+b_{2}+b_{3}-1)/3}\exp\left(  -3z^{1/3}\right)
.
\end{equation}
one can easily show that $P_{\text{Superstatistical}}(y)$\ is reduced to the
Porter-Thomas distribution as the parameter $\alpha\rightarrow\infty$.

Several independent results with different models, have suggested that
transition strengths in a chaotic system follow a $\chi^{2}$ distribution
\begin{equation}
P_{\chi^{2}}(y,\nu)=\frac{1}{2^{\nu/2}\Gamma\left(  \frac{\nu}{2}\right)
}y^{\nu/2-1}e^{-\nu y},
\end{equation}
with $\nu=1$ (porter-Thomas) degrees of freedom, the transition strengths in a
less chaotic system a $\chi^{2}$ distribution with a number of degrees of
freedom less than one. Alhassid and Levine \cite{levine} introduced this
distribution using maximum-entropy arguments. Several studies of
electromagnetic transition intensities in nuclei have also been performed
\cite{alhassid,adams,adamsx,barbosa,shriner,hamoudi}; each of these has
suggested that the $\chi^{2}$ distribution with $\nu<1$ is appropriate for
relatively regular systems.

Experimental data for transition intensities range over several orders of
magnitude. It is often more convenient to consider a logarithmic variable as
the argument. The probability density function in terms of log$_{a}y$ is%
\begin{equation}
F(\text{log}_{a}y)=y\ln a~F(y).
\end{equation}
We compare in Fig. 1 the evolution of $P_{\text{Superstatistical}}($%
log$_{10}y)$ during the transition from chaotic to regular dynamics by varying
the tuning parameter $\alpha$ from $\infty$ (GOE) to $10^{-6}$ (almost
regular) with a corresponding evolution of $P_{\chi^{2}}($log$_{10}y,\nu)$
where $\nu$ varies between 1 and 0.1. Of special interest is the fact that the
maximum of $P_{\chi^{2}}($log$_{10}y,\nu)$ occurs at log$_{10}y=0$ for any
value of $\nu$. This property does not hold for $P_{\text{Superstatistical}}%
($log$_{10}y)$. The peak of superstatistical distribution for less chaotic
systems occurs at log$_{10}y<0$ and moves towards lower values as the
parameter $\alpha$ decreases. We show in the next section that this is indeed
the behavior of physical systems.

\section{DATA ANALYSIS}

The purpose of this section is to show that the proposed superstatistical
distribution succeeds in the situations where the $\chi^{2}$\ distribution
fails. We show this by using results from two works, which examine the effect
of a transition from chaos to integrability on gamma-ray reduced transition probabilities.

The first work is done by Hamoudi, Nazmitdinov and Alhassid \cite{hamoudi}.
They calculated the electric quadrupole (E2) and magnetic dipole (M1)
transition intensities among the isospin $T=0,1$ states of nuclei with mass
number 60. They applied the interacting shell model with realistic interaction
for $pf$ shell nuclei with a $^{56}$Ni core. It is found that the $B$(E2)
transitions are well described by a GOE (Porter-Thomas distribution). However,
the statistics for the $B$(M1) transitions is sensitive to $T_{z}$. The M2
transition operator consists of an isoscalar and isovector components. The
$T_{z}=1$ nuclei, in which both components contribute, exhibit a Porter-Thomas
distribution. In the meanwhile, a significant deviation from the GOE
statistics for the $T_{z}=0$ nuclei, where the matrix elements are purely
isoscalar and relatively weak \cite{brussard}.

We analyzed the reduced M1 transition intensities for both the $T_{z}=1$
$^{60}$Co\ nuclei and $T_{z}=0$ $^{60}$Zn calculated by Hamoudi et al.
\cite{hamoudi} nuclei using the superstatistical transition intensity
distribution in Eq. (14). These authors sampled a large number of matrix
elements for each transition operator,which is equal to $56^{2}=3136$ and
$66^{2}=4356$. Figure 2 compares the results of calculations using Eq. (14)
with the numerical results of Hamoudi et al. \cite{hamoudi} as well as the
"best-fit" $\chi^{2}$\ distribution deduced by these authors. The figure
clearly shows the advantage of the superstatistical distribution proposed here
over the $\chi^{2}$ distribution, at least for this numerical experiment.

The second work that we consider here is that of Adams, Mitchell and Shriner
\cite{adamsx}. They collected approximately 1500 experimental reduced
electromagnetic transition strengths between the excited states of the nucleus
$^{26}$Al. Their data involve levels with isospin $T$ = 0 and $T$ = 1 between
the ground state and the excitation energy of 8.067 MeV. Figure 3 compares
these experimental data with results of calculations using Eq. (14) with
$\alpha=1.24$ as well as the "best-fit" $\chi^{2}$\ distribution with a
parameter $\nu$ slightly greater than 1. The figure again shows the advantage
of the superstatistical distribution over the $\chi^{2}$ distribution,
although the agreement with the data is not as good as in the cases shown in
Fig. 2. The experimental histogram is mostly higher than the theoretical
curves especially in the peak region, although the data was normalized to to
0.83 in order to approximately take care of the upper and lower detection
thresholds. The percentage of undetected transitions may have been
underestimated because its estimation was based on the Porter-Thomas distribution.

\section{CONCLUSION}

The eigenstates of a chaotic system are extended and cover the whole domain of
classically permitted motion randomly, but uniformly. They overlap
substantially, as manifested by level repulsion. There are no preferred
eigenstate; the states are statistically equivalent. As a result, the matrix
elements of transition operators in any basis are independent and have a
Gaussian distribution, which leads to the Porter-Thomas distribution for
reduced transition intensities. Coming out of the chaotic phase, the extended
eigenstates become less and less homogeneous in space. Different eigenstates
become localized in different places and the matrix elements that couple
different pairs are no more statistically equal. The matrix elements will no
more have the same variance; one has to allow each of them to have its own
variance. But this will dramatically increase the number of parameters of the
theory. The proposed superstatistical approach solves this problem by treating
all of the matrix elements as having a common variance, not fixed but
fluctuating. One then expresses the probability density of transition
intensities as an average of Porter-Thomas distributions with different mean
intensities. The principle of maximum entropy is used to estimate the
inverse-mean-intensity distribution. The resulting transition-intensity
distribution is found to agree with the results of shell model calculation as
well as with experimental data better that the $\chi^{2}$ distribution, which
is often used for this purpose.

\pagebreak

\bigskip{\LARGE Figure Caption}

Figure 1. (Color on line) Evolution of the superstatistical distribution
$P_{\text{Superstatistical}}($log$_{10}y)$ and the $\chi^{2}$distirbutions
$P_{\chi^{2}}(log_{10}y,\nu)$\ during the transition from chaotic to regular
dynamics. The solid curves, labeled as PT, refer to the Porter-Thomas distribution.

Fig. 2. (Color on line) Nuclear shell-model M1 transition intensities in
$A=60,$ calculated by Hamoudi et al. \cite{hamoudi}, (histograms) compared
with the superstatistical distribution (14) with parameters $\alpha
=1.499,~0.064~$and $0.030$, respectively (solid curves) and the $\chi^{2}%
$distribution (17) with parameters $\nu=1,~0.64~$and $0.34$, respectively
(dashed curves).

Fig. 3. (Color on line) The distribution of experimental reduced transition
probabilities in 26Al from Ref. \cite{adams} (histogram) compared with the
superstatistical distribution (14) with parameter $\alpha=1.24$ (solid curve)
and the $\chi^{2}$distribution (17) with parameters $\nu=1.04$ (dashed curve).

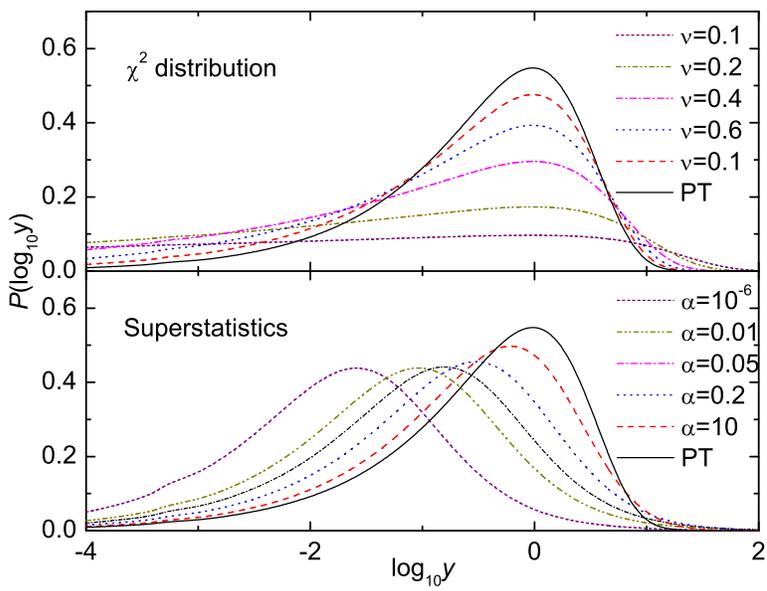
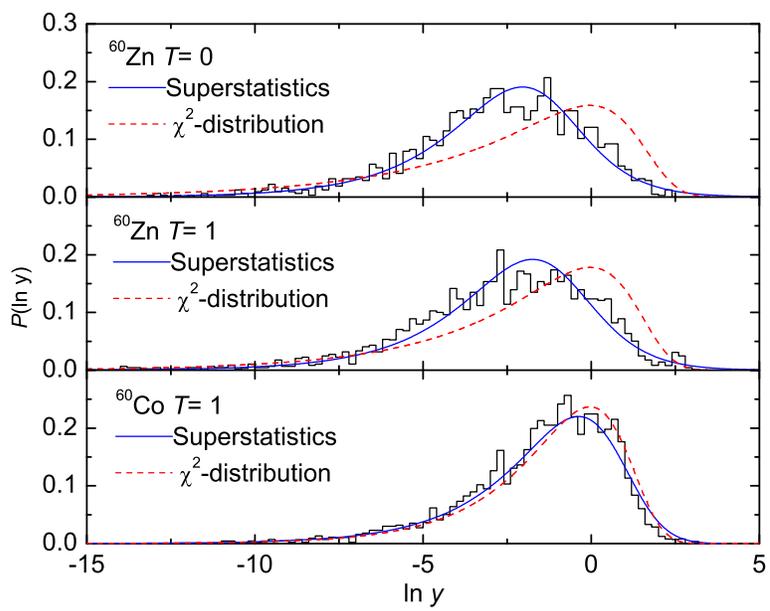
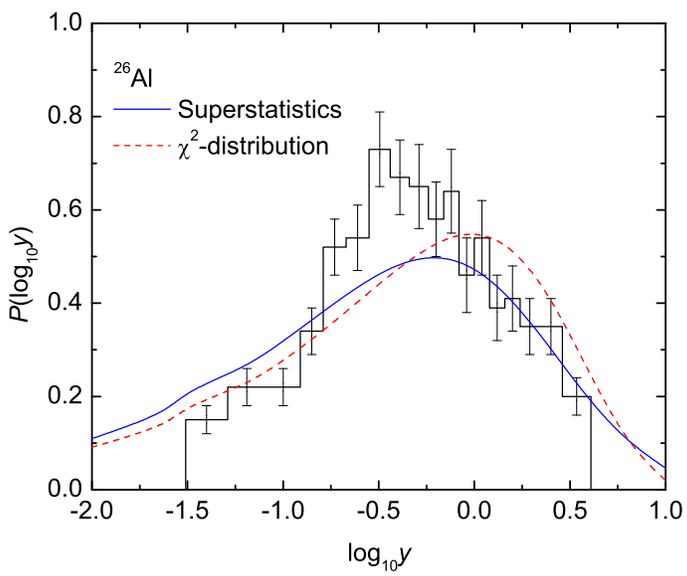
\end{document}